\documentclass{article}

\pdfoutput=1
\usepackage{hyperref}
\hypersetup{
  pdfinfo={
    Title={CT reconstruction from Biplanar DRRs with GA-GANs},
    Author={Ashish Sinha},
    Subject={If you want to put something here, do so},
    Keywords={Computer Vision, GAN, Medical Imaging}
  }
}
\usepackage{pdfpages}

\usepackage[final]{neurips_2019}

\usepackage{lipsum}
\usepackage{graphicx}
\usepackage[utf8]{inputenc} 
\usepackage[T1]{fontenc}    
\usepackage{hyperref}       
\usepackage{url}            
\usepackage{booktabs}       
\usepackage{amsfonts}       
\usepackage{nicefrac}       
\usepackage{microtype}      

\title{GA-GAN: CT reconstruction from Biplanar DRRs using GAN with Guided Attention}

\author{
  Ashish Sinha\thanks{The author was an intern at Preferred Networks, Japan when this work was done.} \\
  Indian Institute of Technology Roorkee\\
  Roorkee, India\\
  \texttt{asinha@mt.iitr.ac.in} \\
   \And
   Yohei Sugawara \\
    Preferred Networks\\
   Tokyo, Japan \\
  \texttt{suga@preferred.jp} \\
   \AND
   Yuichiro Hirano \\
   Preferred Networks \\
   Tokyo, Japan \\
   \texttt{hirano@preferred.jp} \\
}

\begin{document}

\maketitle

\begin{abstract}
  This work investigates the use of guided attention in the reconstruction of CT volumes from biplanar DRRs. We try to improve the visual image quality of the CT reconstruction using Guided Attention based GANs (GA-GAN). We also consider the use of Vector Quantization (VQ) for the CT reconstruction so that the memory usage can be reduced, maintaining the same visual image quality. To the best of our knowledge no work has been done before that explores the Vector Quantization for this purpose. Although our findings show that our approaches outperform the previous works, still there is a lot of room for improvement. 
\end{abstract}

\section{Introduction}

X-Rays have diverse applications in clinical domain. It enables us to see the internal organs of the human body and diagnose the changes like the fractures in the bones. But, projecting all the tissues to 2D space, leads to overlap among images. While bones are clearly visible, the soft tissues are difficult to distinguish. Computed Tomography (CT) overcomes this issue. It helps to visualize the internal organs in a 3D space from a set of X-rays(around 100 - 200) obtained by collecting the photons in full rotation around the body. But getting a CT scan exposes the patient to a high dose of radiation, which may increase the risks of cancer. Additionally, a CT scanner is expensive and bulky, hence not easily accessible in the developing countries or rural areas where transportation is a serious issue.



In this work, unlike the previous methods [\hyperref[sec:1]{1}],[\hyperref[sec:2]{2}], which rely on the use of X-Rays, we use Digitally Reconstructed Radiographs (DRRs). The main motive behind using DRRs, is that an annotated dataset of X-Rays and CT scans are difficult to get. Secondly, the DRRs are of much higher quality as compared to X-Rays and they have a future scope of better image management and DRRs make simulation easier also.

We propose a guided attention [\hyperref[sec:4]{4}] based approach to reconstruct CT volumes from biplanar DRRs that are captured from two orthogonal view planes namely frontal and lateral. It is a difficult problem to match a CT volume to the same input DRR once they are projected onto 2D planes. However, it has been shown that machine learning methods like deep learning are able to learn the mapping between 3D CT volumes and 2D input X-Rays [\hyperref[sec:1]{1}]. 

Attention based GAN approaches have been used to generate natural images, yet using self-attention in the generator makes the training of GAN highly unstable. However, to the best of our knowledge attention based GAN or Vector Quantization(VQ) method have not been used in cross-modality image transfer in medical imaging. In this work, we propose the use of spatial and channel based attention modules in the generator to enhance the visual image quality of reconstructed CT volumes. We also propose the VQ method to generate CT volumes, which is the first of its kind where the encoder is 2D, and the embeddings extracted are 3D to be passed though a 3D decoder and trained in an end-to-end manner.


\begin{figure}
    \centering
    \includegraphics[scale=0.3]{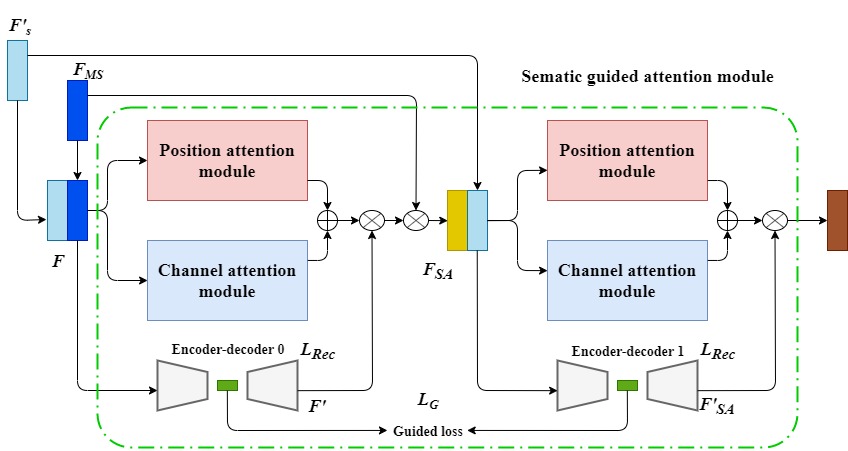}
    \caption{Guided Attention Module  [\hyperref[sec:4]{4}][\hyperref[sec:7]{7}]}
    \label{fig:guided}
\end{figure}
\begin{figure}
    \centering
    \includegraphics[scale=0.25]{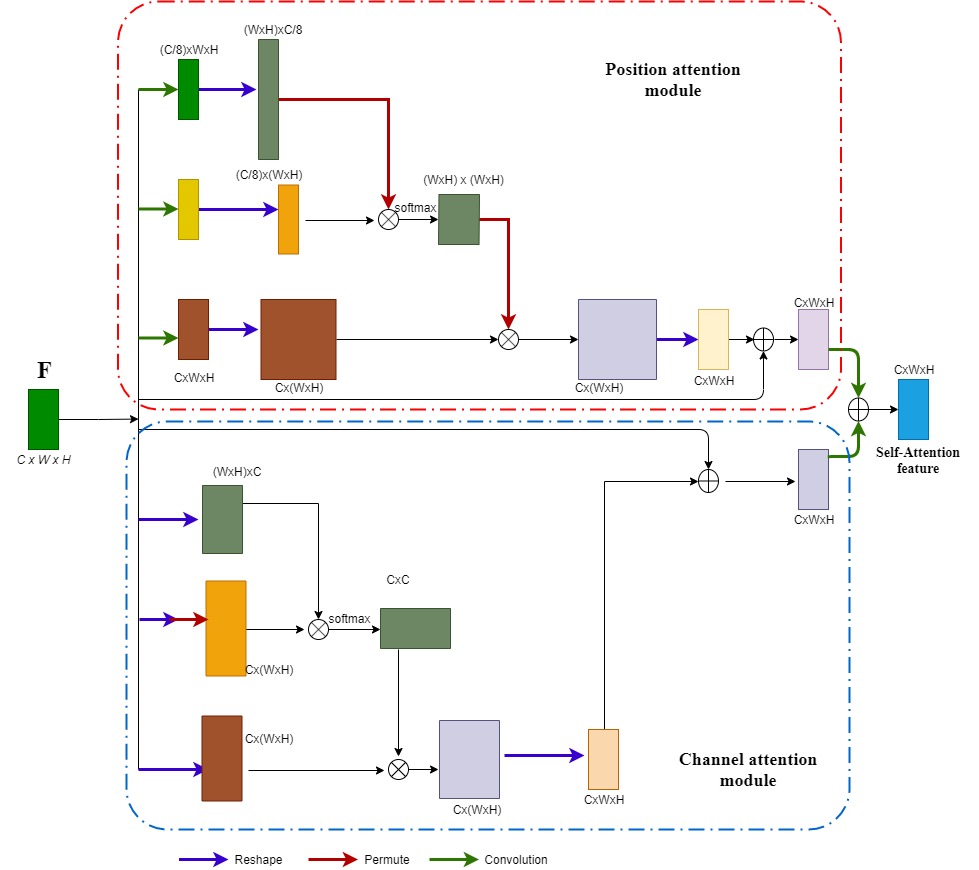}
    \caption{Design of PAM and CAM  [\hyperref[sec:4]{4}][\hyperref[sec:7]{7}]}
    \label{fig:pam}
\end{figure}

  


\section{Methodology}

\subsection{Guided Attention Modules}

Due to the diversity in the structures in medical imaging, traditional CNNs result in the generation of local feature representations and the long-range contextual information is not properly encoded. Even though skip connections are used in the model architecture between the encoder and the decoder, we aim to encode better feature representation before feeding to the decoder. To tackle this problem, we experiment with attention mechanisms to build better association between features. First, the global context is captured by using a multi-scale strategy. Then learned features are fed to the guided attention blocks, which are composed of the spatial and channel attention modules and a semantic attention module. 

Multi-scale features are known to be useful in computer vision problems even before the deep learning era [\hyperref[sec:3]{3}]. Inspired by this work, we make use of learned features at multiple scales, which help to encode both local and global features. Features at different scales are upsampled to a common resolution by bilinear interpolation, then they are concatenated to form a common feature map $F_{MS}$. This $F_{MS}$ is then concatenated with the feature map at last scale to get $F$, and fed into the guided attention module. $F$ is used by both the spatial and channel attention module [\hyperref[sec:4]{4}][\hyperref[sec:7]{7}]. Spatial attention module helps to learn wider and richer contextual representation. Channel attention module helps to learn better feature representation by learning the dependencies between the channel maps. 


Apart from this, we integrate an encoder-decoder network in parallel, that compresses the input features $F$ into a compacted representation in the latent space. The goal is that the class information can be embedded in the second spatial-channel attention module by forcing the semantic representation of both encoder-decoders to be close. Specifically, the feature maps reconstructed in the first encoder-decoder are combined with self attention features generated by the first attention module through a matrix-multiplication to generate $F_{SA}$. Additionally to ensure that the reconstructed feature maps correspond to the features at the input of the spatial and channel attention module, the output of the encoders are forced to be close to their input by using a reconstruction loss. 

\subsection{Discrete Representation Learning}

In this method, we propose to learn the discrete representation of the input DRR image by incorporating ideas from vector quantisation (VQ) [\hyperref[sec:5]{5}]. In this work, we propose an end-to-end model for generating CT volumes by encoding the DRR image into a discrete latent space using a 2D encoder and the sampled 3D embeddings are passed to the 3D decoder. The output of the decoder of the frontal and lateral branches are averaged and upsampled-convoluted thrice to generate the CT. The outputs are averaged by making an assumption that the frontal and lateral DRR images were captured with a negligible time difference. 

\begin{figure}[h]
    \centering
    \includegraphics[scale=0.4]{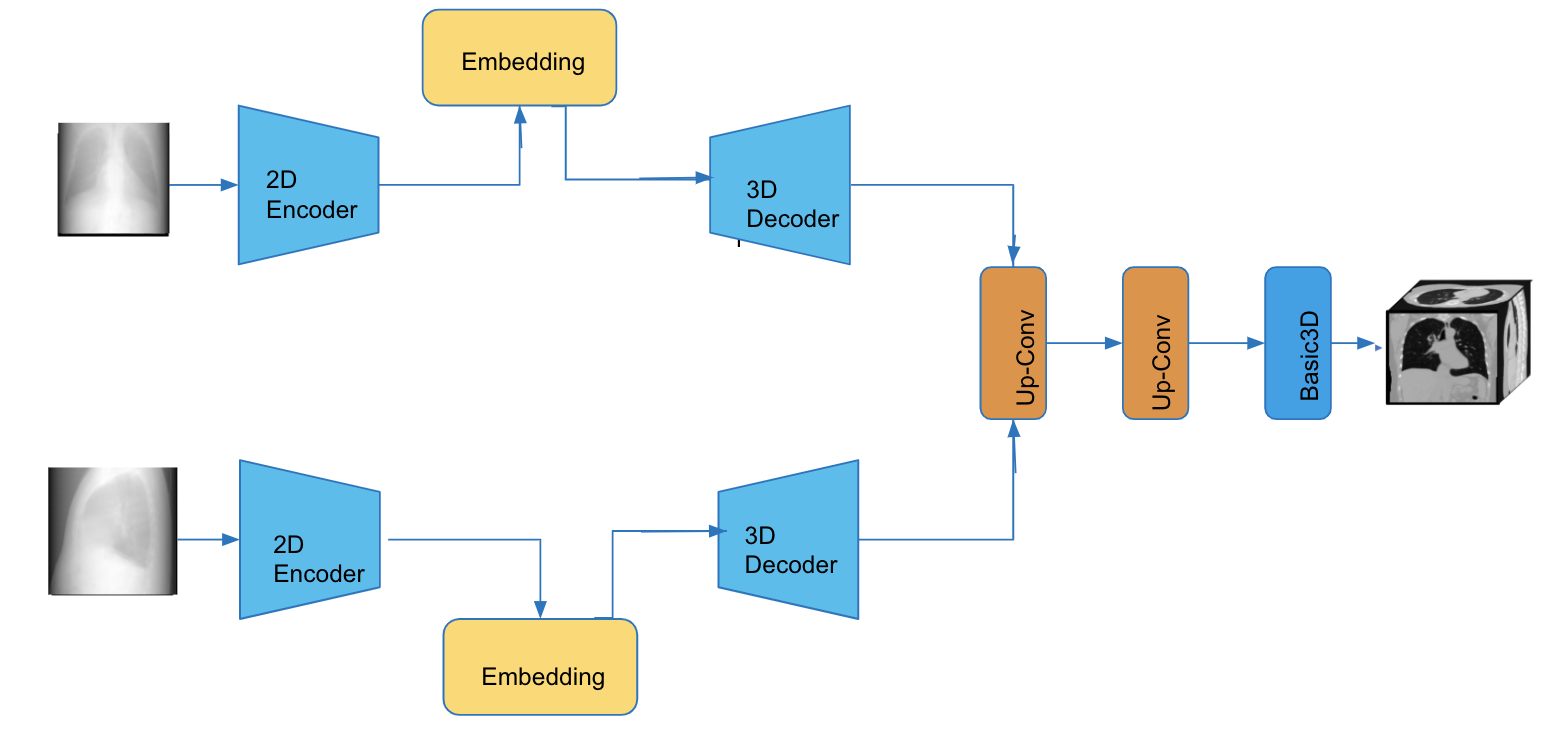}
    \caption{Design of VQ based Generator, the description of components [\hyperref[sec:5]{5}]}
    \label{fig:vq}
\end{figure}

\begin{figure}
    \centering
    \includegraphics[scale=0.3]{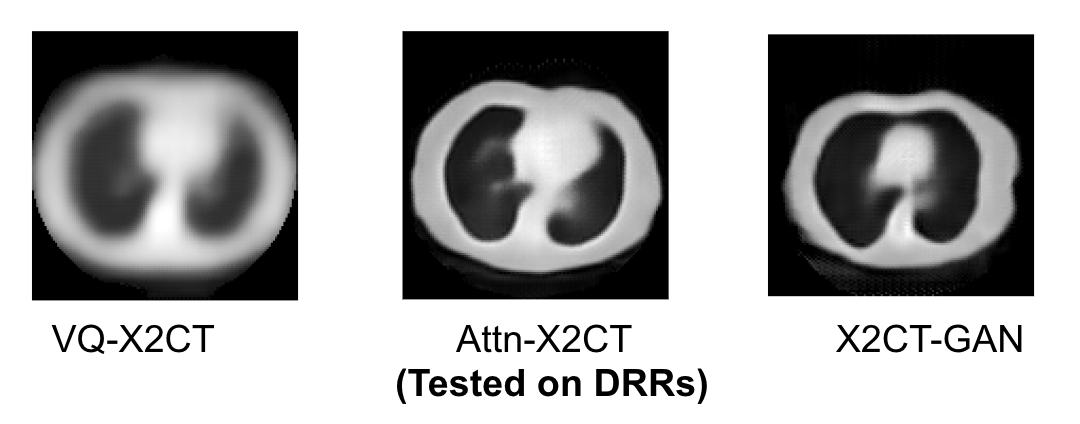}
    \caption{Reconstructed CT samples from DRR input}
    \label{fig:drr}
\end{figure}


    



Since, the output representation of the encoder and the decoder share a different D dimensional space, the forward computation is of the encoder output to the decoder input is simple, but for the backward propagation the output of the decoder is reshaped into 2D space. 

The various components of the VQ based GAN are trained with the loss function similar to $equ. 3$ in [\hyperref[sec:5]{5}]. Additionally, we employ the reconstruction loss, projection loss and a LSGAN loss [\hyperref[sec:8]{8}], as mentioned in [\hyperref[sec:2]{2}]. In our experiments, we found that using a perceptual loss [\hyperref[sec:6]{6}] helps to stabilize the training.

\section{Evaluation}
\subsection{Experimental Setup}
\textbf{Dataset}: We used the LUNA16 dataset for conducting our experiments which contains the 3D CT volumes. The dataset has 10 subsets. We use 9 subsets for the training set and keep 1 subset for validating the results. Each subject contains around 88 scans. 
The input shape is 128x128 for DRR image, and 128x128x128 for the CT voxel. The generated samples are of the shape 128x128x128. 

\textbf{Experiments}: In the experiments, we observed that training the models for a total of 100 epochs is sufficient to reach convergence. We observe that using Adam optimizer for generator and SGD for the discriminator prevents gan failure. We use a learning rate of {$2e-4$} with a weight decay of $1e-6$. We use a exponential decay for learning rate after 50 epochs. 


\begin{table}[h]
  \caption{Quantitative Results}
  \label{results}
  \centering
  \begin{tabular}{lll}
    \toprule
    \cmidrule(r){1-2}
    Method     & PSNR(dB)     & SSIM  \\
    \midrule
    2DCNN & 23.1 & 0.461 \\
    X2CT-GAN & 25.34 & 0.6968 \\
    VQ-GAN (ours) & 23.92 & 0.6259 \\
    \textbf{Attn-GAN (ours)} & \textbf{25.84} & \textbf{0.7156} \\
    \bottomrule
  \end{tabular}
\end{table}

\subsection{Results}
We evaluate the performance of our proposed methods and the baseline model on several widely used metrics like peak signal-to-noise ratio (PSNR) and structural-similarity index (SSIM). PSNR measures the quality of reconstructed signals. SSIM measures the similarity of two images including brightness, contrast and structure. Compared to PSNR, SSIM can measure humans' subjective evaluation better. The results of various models are tabulated in Table \ref{results}. It's visible from $fig$ \ref{fig:drr} that our approach improves the visual quality of reconstructed CT samples, yet there is room for improvement.

\section{References}
\small

\label{sec:1}[1] Henzler, Philipp, et al. "Single-image tomography: 3D volumes from 2D x-rays." arXiv preprint arXiv:1710.04867 (2017).

\label{sec:2}[2] Ying, Xingde, et al. "X2CT-GAN: Reconstructing CT from Biplanar X-Rays with Generative Adversarial Networks." Proceedings of the IEEE Conference on Computer Vision and Pattern Recognition. 2019.

\label{sec:3}[3] Arbelaez, Pablo, et al. "Contour detection and hierarchical image segmentation." IEEE transactions on pattern analysis and machine intelligence 33.5 (2010): 898-916.

\label{sec:4}[4] Sinha, Ashish, and Jose Dolz. "Multi-scale guided attention for medical image segmentation." arXiv preprint arXiv:1906.02849 (2019).

\label{sec:5}[5] van den Oord, Aaron, Oriol Vinyals and Kavukcuoglu Koray. "Neural discrete representation learning." Advances in Neural Information Processing Systems. 2017.

\label{sec:6}[6] Johnson, Justin, Alexandre Alahi, and Li Fei-Fei. "Perceptual losses for real-time style transfer and super-resolution." European conference on computer vision. Springer, Cham, 2016.

\label{sec:7}[7] Fu, Jun, et al. "Dual attention network for scene segmentation." Proceedings of the IEEE Conference on Computer Vision and Pattern Recognition. 2019.

\label{sec:8}[8] Mao, Xudong, et al. "Least squares generative adversarial networks." Proceedings of the IEEE International Conference on Computer Vision. 2017.
\end{document}